\documentstyle[12pt]{JHEP3}

\title{de-Sitter space, (D)pp-Waves and  star-Matrix Model}
 
\author{Harvendra Singh\footnote{Formerly at: Department of Physics, 
Indian Institute of
Technology, Guwahati, India} \\
Theory Division, Saha Institute of Nuclear Physics, \\
1-AF Bidhannagar, Kolkata 700064,  India}

\abstract{ 
We consider  
 de Sitter spacetime solutions 
and the corresponding de Sitter kind of plane-wave solution in M$^*$ theory
which is a $(2+9)$-dimensional theory. 
We attempt to write down  corresponding matrix model which is found 
to have explicit negative energy mass terms as 
well as ghost-like kinetic terms for the matrix valued fields.
 }
\preprint{hep-th/0502011}
\preprint{ \today}
\begin{document}
\def\be{\begin{equation}} \def\ee{\end{equation}}
\def\bea{\begin{eqnarray}} \def\eea{\end{eqnarray}} \def\ba{\begin{array}}
\def\ea{\end{array}} \def\ben{\begin{enumerate}} \def\een{\end{enumerate}}
\def\nab{\bigtriangledown} \def\tpi{\tilde\Phi} \def\nnu{\nonumber}
\def\lll{\label}
\newcommand{\eqn}[1]{(\ref{#1})}
\def\cC{{\cal C}} 
\def\cG{{\cal G}} 
\def\cd{{\cal D}} 
\def\a{\alpha}
\def\b{\beta}
\def\g{\gamma}\def\G{\Gamma}
\def\d{\delta}\def\D{\Delta}
\def\ep{\epsilon}
\def\e{\eta}
\def\z{\zeta}
\def\t{\theta}\def\T{\Theta}
\def\l{\lambda}\def\L{\Lambda}
\def\m{\mu}
\def\f{\phi}\def\F{\Phi}
\def\n{\nu}
\def\p{\psi}\def\P{\Psi}
\def\r{\rho}
\def\s{\sigma}\def\S{\Sigma}
\def\ta{\tau}
\def\x{\chi}
\def\o{\omega}\def\O{\Omega}
\def\k{\kappa}
\def\pa {\partial}
\def\ov{\over}
\def\br{\nonumber\\}
\def\ud{\underline}
\section{Introduction}
The discovery of supersymmetric plane-wave ( or Hpp-wave)
backgrounds in string theory \cite{blau,blau1} has generated lot of
interest in the string community in recent past. The reason behind this
enthusiam has been again the
hope of better understanding AdS/CFT correspondence \cite{maldacena}.
The hint of this fact was demonstated by Blau et. al. \cite{blau1} where
it was shown that plane-wave backgrounds could simply be obtained 
by applying the Penrose 
limits \cite{penrose,gueven} on (anti)de Sitter spacetimes, 
$(A)dS \times Sphere$. The
 plane-wave limits
\cite{blau1} basically make us zoom in onto those null geodesics 
which have a direction along sphere. 
Another fact  is that  string theory in 
pp-wave background becomes exactly solvable theory in 
suitable light-cone gauge \cite{matsaev,matsaev1}.
Therefore, from AdS/CFT correspondence point of view \cite{maldacena}, 
a pp-wave spacetime in bulk has useful consequences for
dual conformal field theories on the boundary \cite{bmn}. 
Under this dictionary called BMN-correspondence \cite{bmn}, 
 type IIB closed string excitations 
in a pp-wave background spacetime are dual to
 conformal field theory operators with 
large $U(1)$ R-charge. Since then there has been much progress
in this field.

In this work we would like to concentrate on the plane-wave geometries
in type II$^*$ string theories \cite{hull}, which admit another kind of 
plane-waves known as de Sitter plane-waves (Dpp-wave) in
\cite{har04}. These Dpp-waves can be obtained by taking pp-wave limits
of corresponding de Sitter spacetime solutions of type II$^*$ theories
\cite{har04}. In this view, Dpp-waves can be as useful for dS/CFT
holography
\cite{witten} as are the supersymmetric Hpp-waves for AdS/CFT holography.  

It was proposed some time back by Hull \cite{hull} that if we consider 
time-like T-duality of usual type II string 
theory with spacetime signature $(1,9)$, then one can study string 
theories 
on arbitrary $(m,n)$ signature spacetimes. This duality along 
time direction somehow takes 
away the significance of the spacetime signature from the string theory, in 
particular the significance of Lorentzian symmetry  $SO(1,9)$.   
However, while implementing time-like duality it requires the assumption 
of a compact 
time coordinate. It then makes the RR fields of the T-dual type II theory 
tachyonic (ghost-like) in nature, i.e., the kinetic terms with negative 
sign  appear in the low energy supergravity  action. These type II
string theories in $(1+9)$ dimensions with  
tachyonic kinetic terms have been designated as II$^*$ string theories 
\cite{hull}.
Nevertheless, these star-theories are maximally supersymmetric
string theories.
We would like to note that the theories with tachyonic matter fields or
with a tachyonic potential have been  favourable 
candidates for cosmology,
see \cite{ashoke} and  references therein. The motivation is that we
should be  able to obtain
de-Sitter spacetime solutions from string theory. The de Sitter spacetimes
naturally appear as solutions in 
II$^*$  theories. The
usual string theory forbids de-Sitter space as a classical
solutions, however see \cite{kachru}
for recent developments.

This paper is organised as follows. In section-2 we first review the
$dS_5\times H^5$
de Sitter solution and corresponding Dpp-wave in type
IIB$^*$ supergravity. 
Then we discuss time-like T-duality map between
Hpp-wave and the Dpp-wave solutions. In section-3 we write down the 
analogous de Sitter plane-wave or D(M)pp-wave solution in $(2+9)$
dimensional
M$^*$ theory. We then construct BMN-like matrix model which contains
negative energy kinetic terms as well as negative energy mass terms for
the bosonic fields. We find that fuzzy
2-sphere solutions still form the ground state of the matrix model.
The conclusions are given in
section-4.

\section{de-Sitter plane-waves (Dpp-wave)}
 
\subsection{The $dS_5\times H^5$ and a Dpp-wave solution}

Consider type IIB$^*$ supergravity \cite{hull}, the action has the kinetic 
terms for all the  
RR-potentials with negative signs. This theory can be obtained by 
implementing T-duality along compact time direction of ordinary type IIA 
theory. 
We shall assume that the II$^*$ theories
are consistent string theories, when considered on a compact time-like
circle, even though they have 
wrong sign kinetic terms of RR potentials in the low energy 
supergravity action. 
A quite remarkable property of these star-theories is that, 
there do exist plenty of de-Sitter spacetime solutions in them, e.g.
$dS_5\times H^5$ background in 
type IIB$^*$ theory \cite{hull2,hull1,liu}. As is
shown in detail
\cite{hull2} \footnote{I am thankful to
Shibaji for bringing this work to my notice and for discussion.}, 
these $dS_5$ vacua can be obtained by taking near
horizon
limits of Euclidean E$4$-brane solutions of IIB$^*$
theory.\footnote{Half-supersymmetric E$n$-branes 
in type II$^*$ theories have spacelike $n$ world-volume directions and
$10-n$ 
transverse directions which include the time coordinate (with Dirichlet 
boundary conditions) \cite{hull2}.}  They  are also maximally
supersymmetric solutions. These E4-brane solutions are  useful for
dS/CFT holography, as are the D3-branes in AdS/CFT holography, for
example.
The corresponding dual super-Yang-Mills theory is defined in
4-Euclidean dimensions.

The de Sitter plane wave solutions can be obtained by considering
Penrose-G\"uven limits 
 of the $dS_5\times H^5$ background. 
These Dpp-wave solutions \cite{har04}
 obtained through Penrose limits are  given by

\bea\lll{newwave}
&& g:= 2dx^{+}dx^{-} +(\mu^2\delta_{ab}x^ax^b)  (dx^{-})^2+ 
\sum_{a=1}^{8}(dx^a)^2\br
&& F_{-1234}=2\sqrt{2} \mu=F_{-5678}
\eea
It can be seen that only nonvanishing 
componant of the Ricci tensor for the Dpp-wave metric \eqn{newwave} is 
$R_{--}=-8 \mu^2 \ ,$ 
which is strictly negative.\footnote{We follow the convention where 
for a de 
Sitter space of radius of curvature $l$, the Ricci tensor is given by 
$R_{\m\n}={(D-1)\over l^2}g_{\m\n}$. Here $D$ is the spacetime 
dimension.} Also note that the coefficient of the  $(dx^{-})^2$ term in 
the Dpp-wave metric is   
 positive definite which is opposite of 
the case of Hpp-wave \cite{blau}. To note Hpp-wave is a maximally
supersymmetric pp-wave
solution in type IIB theory.
The reason for this is that the background \eqn{newwave} satisfies 
the field equations 
of  type IIB$^*$ supergravity where $C_{(4)}$ kinetic term has negative
sign. 

\subsection{ The time-like T-duality}

We briefly discuss time-like duality map for the Dpp-wave solutions.
The time-like T-duality 
map between the fields of two types of string theories has been 
constructed by Hull 
\cite{hull}. Under this map the background fields of type IIB (IIA) string 
theory 
are mapped to the fields of IIA$^*$ (IIB$^*$) theory and vice-versa. We
would like to show that Hpp-wave and the Dpp-wave are mapped to each other
under the time-like T-duality along with space-like T-duality. 
Consider first the supersymmetric Hpp-wave solution of type IIB 
theory \cite{blau}
\bea\lll{Hpp}
&& g:= 2dx^{+}dx^{-} +(H_{ab}x^ax^b)  (dx^{+})^2+ 
\sum_{a=1}^{8}(dx^a)^2\br
&& F_{(5)}= 2\sqrt{2}\mu~ dx^{+}(
dx^1dx^2dx^3dx^4
+ dx^5dx^6dx^7dx^8)
\eea
with the matrix $H_{ab}=-
\mu^2\delta_{ab}$. The Ricci tensor for Hpp-wave is
$R_{++}=8\mu^2  \ , $
which is positive  unlike the Dpp-wave. In \eqn{Hpp} we can consider
$x^{+}$ as time coordinate while $x^{-}$ acts like a
space-like direction. In this sense the 5-form flux is electric type. We
can compactify
$x^{+}$ coordinate  and make
a time-like duality \cite{hull}. Using the duality relations, 
the background obtained as a result is  extended string solution
in type IIA$^*$ theory \cite{har04}.
In subsequent step we  make a space-like T-duality by compactifying
the space-like string direction $x^{-}$ of the string solution. That 
gives us 
a type IIB$^*$ plane wave solution \cite{har04}
\bea\lll{newwave1}
&& g:= 2dx^{+}dx^{-} +(\mu^2\delta_{ab}x^ax^b)  (dx^{-})^2+ 
\sum_{a=1}^{8}(dx^a)^2\br
&& F_{(5)}=2\sqrt{2}\mu dx^{-}
(dx^1dx^2dx^3dx^4+dx^5dx^6dx^7dx^8),
\eea
which is nothing but the Dpp-wave as 
in \eqn{newwave}. Since in \eqn{newwave1} the coordinate
$x^{-}$ behaves like a space direction, the flux $F_{(5)}$ is magnetic 
type. Hence the roles of $x^{+}$ and $x^{-}$ have got exhanged under the 
above duality 
map which includes one time-like T-duality. However this is not of much 
significance because $x\to -x$ will exchange 
$x_{-}\leftrightarrow x_{+}$. 
The Dpp-waves are maximally supersymmetric as they are obtained from 
Penrose limits of the corresponding supersymmetric $dS_5\times H^5$
solutions \cite{har04}. Even otherwise the T-dualities preserve
supersymmetry of the given background.

Thus we see that starting from Hpp-wave of IIB string and implementing a  
time-like duality and a space-like duality in succession, we have obtained 
Dpp-wave which is a
solution of the IIB* string theory. 
In the next section we discuss de Sitter plane-wave solution in M$^*$
theory. The motivation is that we should be able to study dS/CFT
holography through Dpp-wave/CFT holography as in the
BMN-model \cite{bmn}.

\section{A D(M)pp-wave background in M$^*$ theory}
As we know M-theory compactification on a square 2-torus, $T^{2}$, 
in the limit 
of shrinking torus size leads to type IIB string theory. 
Similarly, 
M$^*$-theory compactification on a Lorentzian torus, $T^{1,1}$, in the 
limit 
of torus shrinking to zero area leads to type IIB$^*$ strings 
\cite{hull}. So 
it is natural to expect that the Dpp-wave like solutions in M$^*$-theory. 
The M$^*$ theory has $(2,9)$
spacetime signature. The supergravity action has the
kinetic energy 
term for the 3-form potential with 
negative sign and is
given by
\cite{hull} 
\begin{eqnarray} 
S&=&\int \bigg[ \left\{
R~^{\ast}1 +{1\over2} G_{(4)}
~^{\ast} G_{(4)} \right\}
+{4\ov3} d C_{(3)} d C_{(3)}  C_{(3)} \bigg]\ ,
\lll{Mstar}
\end{eqnarray}
The field 
strength is defined through $G_{(4)}=dC_{3}$. 
It is straight forward and obvious that
there exists a Freund-Rubin type de-Sitter solution $dS_{(1,3)}\times
H^{(6,1)}$ in M$^*$ 
theory. 
This solution is rather interesting as it has four-dimensional de Sitter
sub-spacetime and a seven-dimensional hyperbolic spacetime. 
The pp-wave limit  of this de Sitter solution
provides us with the following 11-dimensional plane-wave solution
\cite{blau1}
\bea\lll{Hpp1}
&& g:= 2dx^{+}dx^{-} +H(x)  (dx^{-})^2+ \eta_{ab} dx^adx^b\br
&&G_{(4)}=\mu dx^{-}dx^1dx^2dx^3 ,
\eea
 where metric $\eta_{ab}={\rm diag}(+,\cdots,+,-)$ is of indefinite 
signature involving one time like transverse direction  and 
the function $$H= 
{\mu^2\over9}[(x^1)^2+\cdots+(x^3)^2]+{\mu^2\over36}[(x^4)^2+\cdots-(x^9)^2].$$ 
We will distinctly call the solution \eqn{Hpp1} as DMpp-wave.
These  DMpp-waves are endowed with noncompact rotational isometry 
group $SO(3)\times 
SO(5,1)$. This is quite unlike the usual supersymmetric (M)pp-wave of 
Matrix-theory 
\cite{bmn}, which 
has compact rotational isometry group $SO(3)\times SO(6)$ in the
transverse directions. 
Also, $H(x)$ is indefinite here and specifically time dependent. We
must note that  time-independent DMpp-waves can also be obtained,
but we want to discuss most general case here.

\subsection{BMN$^*$ matrix model}
Having obtained the eleven-dimensional DMpp-wave background 
in \eqn{Hpp1} it is imperative to discuss the BMN type matrix model 
\cite{bmn} 
for this background.  
The BMN model \cite{bmn} admits fuzzy 2-spheres and 5-branes as
supersymmetric
solutions. Let us assume $x^{+}\sim 
x^{+} +2\pi R $, being periodic over a circle of radius and 
$R\equiv R_{BFSS}$ \cite{bfss}.    
Then the dynamics of the theory in the momentum sector $p^{-}=p_{+}=N/R$ 
will be given by $U(N)$ matrix model. This can be understood in the 
following way.

 As we know, M$^*$ theory is considered as the (time-like)
decompactification of  type IIA$^*$ string theory in the strong coupling 
limit. At the same time analogue of D0-branes are the E1-branes
in IIA$^*$ theory \cite{hull2}. These supersymmetric $N$ E1-branes then
will describe BFSS-like $U(N)$ matrix model   
\bea 
&&S'_{BFSS}=\int dt Tr \bigg[{1\ov2R}(D_t x^a)^2 -{R\ov4} [x^a,x^b]^2 
+{\rm fermionic ~terms}\bigg]
\label{m1model}
\eea
where indices $a,b=1,2,\cdots,9$ and 
time $t$ is identified with $x^{-}$.\footnote{ The time $t$ in this
section must be viewed as Euclidean time.} 
Indices $a,b$ are contracted 
with Minkowski metric $\eta_{ab}$. Comparing \eqn{m1model} to the BFSS
matrix 
model \cite{bfss}, the kinetic term for field $x^9$ are
with negative signs in \eqn{m1model}, i.e. those are
{\it ghost} like in nature. To note, the  action \eqn{m1model} has been
obtained by 
placing the Minkowskian metric $\eta_{ab}$ in the internal 
(rotational) space of the fields. Due to this the model has a noncompact
rotational group symmetry $SO(1,8)$. The noncompact nature of isometry 
group $SO(1,8)$ is due to the fact that E1-branes, which give rise to 
matrix model dynamics at strong coupling, have one time direction and
eight space
directions with 
Dirichlet 
boundary conditions.\footnote{The E1-branes though have tachyonic mass but
become
massless in the strong coupling limit, quite similar to the D0-branes
becoming massless in matrix-model picture \cite{hull2}.}
The easiest way to ensure that all the signs are
correct in the matrix model is to obtain it by compactifying 
ten-dimensional
$SU(N)$  Super-Yang-Mills theory on $(1,8)$ torus.

Then corresponding to the DMpp-wave background, BMN like action can
be written in analogous manner and we propose 
\bea &&S=S'_{BFSS}+S_{mass} 
\br 
&&S_{mass}=\int dt~\bigg[ 
{1\ov2R}Tr[{\mu^2\over9}(x_1^2+\cdots+x_3^2)+
{\mu^2\ov36}(x_4^2+\cdots-x_9^2)]-{i\mu\ov3} 
\ep_{rls}Tr(x^rx^lx^s) 
\br&&+{\rm 
fermionic~ terms}\bigg] 
\label{mmodel}
\eea
where indices $a,b=1,\cdots,9$; while $r,l=1,2,3$.  
Comparing \eqn{mmodel} to the BMN matrix 
model \cite{bmn}, the kinetic term for the field $x^9$ and its mass term   
appear with negative signs in \eqn{mmodel}, i.e. those are
{\it tachyonic} in nature. The above action has been obtained by 
placing the Minkowskian metric $\eta_{ab}$ in the internal 
isometry (rotations) space of the 
fields $x^a$ in the BMN model and simultaneously changing the signs 
of all kinetic terms. However, we must note that it is now an Euclidean
matrix model. The Myers term remains
the same as in BMN model 
because the $G_{(4)}$ flux remains unchanged. The mass terms break the
symmetry from $SO(1,8)$ to $SO(3)\times SO(5,1)$. If we set $\mu=0$ in
the 
above BMN$^*$ matrix model \eqn{mmodel} we  obtain  BFSS-like matrix
model \eqn{m1model},
but one of the 
scalar field $x^9$ will be tachyonic owing to the noncompactness of the 
transverse isometry group. We have already mentioned, this is due to fact
that we have E1-branes with time-like coordinate in the transverse space. 
\subsection{Solutions} 
We do find that the {\it fuzzy} 
2-sphere is {\it still} a solution of the above matrix model \eqn{mmodel}. 
If we take
$$[x_r,x_l]=i {\mu\over3R}\epsilon_{rls}x_s$$ with 
$\dot x^a=0$, $ 
x^4=0=\cdots=x^9$, 
the classical equations of motion are satisfied.
It will correspond to fuzzy sphere of physical radius
in 1,2,3 space-like directions given by
\be
r\sim 2\pi \sqrt{Tr\sum_r x_r^2\over N} \sim \pi {\mu N\over 3R}.
\ee
We cannot say anything about the supersymmetries preserved by this
solutions. But, we expect them to be supersymmetric and stable.

The other interesting 
solutions are the {\it fuzzy} 2-torus type \cite{bigatti}. Consider a pair 
of $U(2)$ matrices $x^4,x^5$  we have a $noncommuting$ solution  
\bea
&& x^4x^5=x^5x^4 e^{i\theta},~~~\theta\equiv{2\pi\over N}=\pi ~ ({\rm 
since~} 
N=2)  \br
&& (x^4)^2=(x^5)^2=1, ~~~(x^4)^\dagger x^4=1=(x^5)^\dagger x^5
\label{ftorus} \eea
with $\dot x^a=0$, 
 $x^1=x^2=x^3=0, x^7=0=\cdots=x^9$, provided mass parameter 
${\mu\over 2R}=12$. Such static 
solutions are possible for any 
pair of indices $a,b$ from the set of indices $4,5,\cdots,9$. 
We can't again say definitely what would be the supersymmetry. 

The time-dependent $commuting$ configurations of the type
\bea
&&
 x^1 (t)= e^{-{2\mu\over 3R}t} x^1(0),~~x^2 (t)= e^{-{2\mu\over 3R}t} 
x^2(0)\br
&& [x^1(0),x^2(0)]=0
\eea
are also solutions. Similar time-dependent decaying solutions exist for 
any other pair of 
indices $a,b$ from $1,2,3$ or $4,\cdots,9$.
   
\section{Conclusions}

We have studied de Sitter plane wave (or DMpp-wave) solutions in M$^*$
theory in (2,9) spacetime dimensions. These DMpp-waves are obtainable by
employing the 
Penrose-G\"uven limits on the $dS_{(1,3)}\times H^{(6,1)}$ background of
M$^*$ theory.
We find these DMpp-waves are quite analogous to maximally supersymmetric 
pp-wave geometry in  M-theory, but have $SO(3)\times SO(5,1)$ 
non-compact rotational isometries. We have also written down BMN-like
matrix model 
action for the DMpp-wave background. The matrix model allows the
kinetic terms of fields with negative signs along with tachyonic mass 
terms. It is an expected feature because original low energy supergavity
action for 
M$^*$ theory has all these qualities \cite{hull}. Nevertheless it has been
interesting to observe that the new BMN$^*$ matrix model with tachyonic 
fields admits fuzzy 2-spheres as vacuum solutions which must be
supersymmetric. We note that BMN
model \cite{bmn} has supersymmetric (stable) fuzzy 2-sphere vaccum
solutions.

We would like to summarise by saying that it will  be interesting to study
further the BMN$^*$ matrix model which quite nicely encodes the 
string (membrane) dynamics in the de Sitter backgrounds. Further we think
that it will be worth while to understand the nature of
Dpp-wave/CFT correspondence in
type II$^*$ string theories and the strong coupling picture
through BMN$^*$ matrix model, and hope that we may be able to
understand the exact nature of dS/CFT correspondence \cite{witten}.

\leftline{\bf Acknowledgements}
I am grateful to Shibaji Roy for carefully reading the manuscript and for
the comments.
I am also thankful to the organisers of International Workshop on String 
Theory (ISM04) in 
Khajuraho for the invitation to present this work. The major part of this
work has been carried out at Indian Institute of Technology, Guwahati.      
  

\end{document}